\documentclass{phc-proc4-auth}
\usepackage{graphicx,dcolumn,bm,amssymb}
\begin{document}

\begin{frontmatter}
%%%%%%%%%%%%%%%%%%%%%%%%%%%% TITLE
\title{Field-induced quantum critical point in CeCoIn$_5$}

%%%%%%%%%%%%%%%%%%%%%%%%%%%% AUTHORS
\author[TOR]{Johnpierre~Paglione},
\author[TOR]{M.~A.~Tanatar\thanksref{MAT}},
\author[TOR]{D.~G.~Hawthorn},
\author[TOR]{Etienne~Boaknin},
\author[TOR]{R.~W.~Hill},
\author[TOR]{F.~Ronning},
\author[TOR]{M.~Sutherland},
\author[TOR]{Louis~Taillefer\corauthref{COR}\thanksref{LT}},
\author[AMES]{C.~Petrovic\thanksref{BNL}},
\author[AMES]{P.~C.~Canfield}

\corauth[COR]{Corresponding author: Louis.Taillefer@physique.usherb.ca}
\address[TOR]{Department of Physics, University of Toronto, Toronto, Ontario, Canada M5S 1A7}
\address[AMES]{Ames Laboratory and Department of Physics and Astronomy, Iowa State University,
Ames, Iowa 50011}
\thanks[MAT]{Permanent address: Inst. Surf. Chem., N.A.S. Ukraine.}
\thanks[LT]{Current address: Department of Physics, University of Sherbrooke, Sherbrooke, Canada
J1K 2R1}
\thanks[BNL]{Current address: Department of Physics, Brookhaven National Laboratory, Upton, New
York 11973}

%%%%%%%%%%%%%%%%%%%%%%%%%%%% ABSTRACT
\begin{abstract}
The resistivity of CeCoIn$_5$ was measured down to 20 mK in magnetic fields of up to 16 T. With
increasing field, we observe a suppression of the non-Fermi liquid behavior, $\rho \sim T$, and the
development of a Fermi liquid state, with its characteristic $\rho=\rho_0+AT^2$ dependence. The
field dependence of the $T^2$ coefficient shows critical behavior with an exponent of $\sim 4/3$.
This is evidence for a new field-induced quantum critical point, occuring in this case at a
critical field which coincides with the superconducting upper critical field $H_{c2}$.
\end{abstract}


\end{frontmatter}

%%%%%%%%%%%%%%%%%%%%%%%%%%%% INTRODUCTION

The recent discovery of a new family of heavy-fermion superconductors with general formula
CeMIn$_5$ (M=Co, Ir, Rh or their solid solutions) has generated much interest \cite{Cedomir}.
CeCoIn$_5$, with ambient pressure superconductivity below $T_c=2.3$~K, is close to a point where
the magnetic state becomes unstable as $T\rightarrow 0$ \cite{QCP_115}. This proximity to a quantum
critical point (QCP) is believed to be responsible for the unusual superconducting and normal state
properties, but not much is known about the QCP itself.

Although long-range magnetic order is not present in CeCoIn$_5$ \cite{uSR}, the close proximity of
this system to antiferromagnetic (AF) order \cite{Kohori} results in an abundance of spin
fluctuations which lead to deviations from Fermi liquid (FL) behavior \cite{Cedomir,NFL}. Since
magnetic fluctuations play an essential role in quantum criticality, the response of this system to
applied magnetic fields is of clear interest. In order to help elucidate the nature of the QCP, we
have measured the low $T$, in-plane resistivity, $\rho$, of single-crystal samples of CeCoIn$_5$ in
transverse fields, $H$, applied parallel to the c-axis.

%%%%%%%%%%%%%%%%%%%%%%%%%%%% RESULTS AND DISCUSSION

\begin{figure}
 \centering
 \includegraphics[totalheight=2.7in]{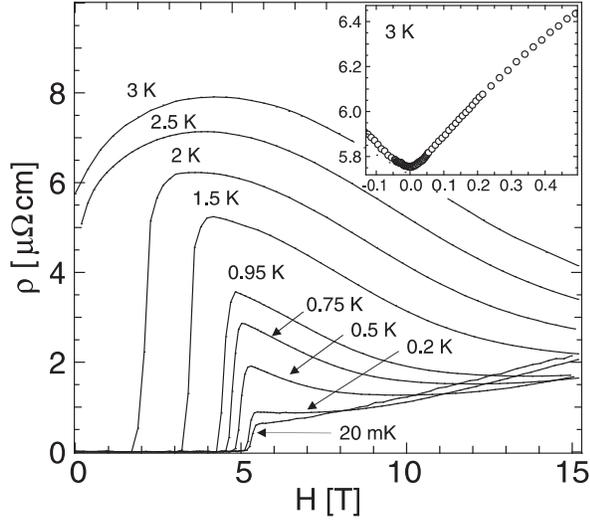}
 \caption{\label{fig:MR} Resistivity of CeCoIn$_5$ from constant-temperature field sweeps, shown as
 isotherms versus applied field. The inset displays a zoom of the anomalous linear
 magnetoresistance observed at low fields in the normal state (3~K).}
\end{figure}

In CeCoIn$_5$, notable magnetoresistance (MR) begins to appear below 30~K: the linear $T$
dependence of $\rho(T)$ observed at $H=0$ suffers a drastic change with increasing $H$, most
pronounced at low $T$. The field dependence of $\rho$, obtained in constant temperature $H$-sweeps,
is plotted in Fig.\ref {fig:MR}. The evolution of $\rho(H)$ with increasing temperature reveals the
development of a {\it crossover} in the sign of MR with increasing $H$ and $T$. This crossover is
traceable to higher temperatures, and is a clear indication of a field-induced change in character
of the spin fluctuations residing in this system \cite{Paglione}. The normal state, low-field MR
(inset of Fig.~\ref{fig:MR}) displays a notable range of {\it linear} field dependence, reproduced
in negative field. This behavior, which is quite different from the usual $H^2$ dependence
observed at such low fields in conventional metals, is not the $H$-linear MR often observed in the
$\omega_c \tau >> 1$ limit \cite{Paglione,Pippard}. Rather, it may indicate the close proximity of
CeCoIn$_5$ to a zero-field QCP \cite{Rosch}.

%%%%%%%%%%%%% figure 1

\begin{figure}
 \centering
 \includegraphics[totalheight=2.5in]{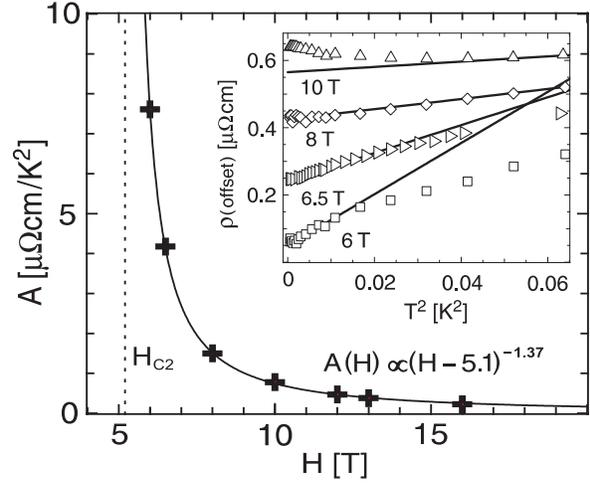}
 \caption{\label{fig:AvsH} Magnetic field dependence of the quadratic-term coefficient $A$ of
 $\rho(T)$. The solid line is a fit of the data points ($\bf +$) to the displayed formula. The
 inset displays the low temperature resistivity of CeCoIn$_5$ plotted vs. $T^2$ for several
 magnetic fields (data offset for clarity). The lines are linear fits to the data, showing the
 evolution of $A$ (slope) with magnetic field.}
\end{figure}

An analysis of the $\rho(T)$ data obtained at high (constant) fields reveals the development of a
clearly distinguishable range of $\rho \sim T^2$ behavior at low $T$. As shown by the linear fits
of $\rho$ vs. $T^2$ in the inset of Fig.~\ref{fig:AvsH}, the range of $T^2$ behavior is small at
lower fields, appearing only below $\sim$100~mK at 6~T. This range, which gradually becomes wider
and more apparent with increasing field \cite{upturn}, extending up to 2.5~K by $H$=16~T, is
identified with the recovery of a FL regime for $H \gtrsim 5$T.  Concomitantly, the slope of the
fitted curves, {\it i.e.} the coefficient $A$ in $\rho = \rho_0 + A T^2$, dramatically decreases
with increasing field. The field dependence of $A$, or $A(H)$ (Fig.~\ref{fig:AvsH}), displays {\it
critical} behavior, best fitted by the function $A\propto (H-H^*)^{-\alpha}$ with parameters
$H^*=5.1\pm0.2$~T and $\alpha=1.37\pm0.1$, indicative of a divergence in the strength of electron-
electron interactions at the critical field $H^*$. This behavior, taken together with the
development of a FL state above $H^*$, is commonly associated with quantum critical behavior,
revealing the existence of a {\it field-induced} QCP in CeCoIn$_5$.

%%%%%%%%%%%%% figure 2

A number of other systems exhibit similar critical behavior in resistivity when approaching some
critical field value associated with a QCP \cite{QCP_other,YbRhSi}. What is unique (and intriguing)
about CeCoIn$_5$ is the fact that $H^* \approx H_{c2}(0)=5.1$~T, which points to the existence of a
QCP coincident with the superconducting transition at $T=0$. However, because the $H_{c2}$
transition in CeCoIn$_5$ is first-order below $\sim 0.7$~K \cite{firstorder}, it is tempting
to propose that the QCP is not associated with the superconducting state, but rather with a $T=0$
transition of magnetic origin. This interpretation is reinforced by the fact that in
antiferromagnetic Ge-doped YbRh$_2$Si$_2$, one also finds $A \propto (H-H*)^{-\alpha}$ with
$\alpha~\sim~4/3$, but where $H^*=H(T_N\rightarrow 0)$ \cite{YbRhSi}.

%%%%%%%%%%%%%%%%%%%%%%%%%%%% CONCLUSIONS

In conclusion, we have identified the anomalous low-temperature evolution of magnetoresistance
in CeCoIn$_5$ with the development of a Fermi liquid regime away from a field-induced quantum
critical point.

%%%%%%%%%%%%%%%%%%%%%%%%%%%% ACKNOWLEDGMENTS

This work was supported by the Canadian Institute for Advanced Research and funded by NSERC.

%%%%%%%%%%%%%%%%%%%%%%%%%%%% BIBLIOGRAPHY


\begin{thebibliography}{00}
\bibitem{upturn} At low temperatures, a small upturn in $\rho(T)$ also starts to develop in fields
above 8~T, but does not affect the $T^2$ analysis and is explained elsewhere \cite{Paglione}.
\end{thebibliography}
\end{document}